\documentclass[twocolumn,preprintnumbers]{revtex4}
\usepackage{graphicx}
\usepackage{dcolumn}
\usepackage{bm}
\usepackage[latin1]{inputenc}

\begin{document}

\title{Rotons in a hybrid Bose-Fermi system}
\author{Ivan A. Shelykh}
\affiliation{Science Institute, University of Iceland, Dunhagi-3, IS-107, Reykjavik,
Iceland}
\affiliation{International Institute for Physics, UFRN - Universidade Federal do Rio
Grande do Norte, Campus Universitario Lagoa Nova, CEP: 59078-970, Natal- RN,
Brazil}
\author{Thomas Taylor and Alexey V. Kavokin}
\affiliation{School of Physics and Astronomy, University of Southampton, Highfield
Southampton, SO171BJ, United Kingtom}
\date{\today }

\begin{abstract}
We calculate the spectrum of elementary excitations in a two dimensional
exciton condensate in the vicinity of a two-dimensional electron gas. We
show that attraction of excitons due to their scattering with free electrons
may lead to formation of a roton minimum. The energy of this mimimum may go
below the ground state energy which manifests breaking of the superfluidity.
The Berezinsky-Kosterlitz-Thouless phase transition temperature decreases
due to the exciton-exciton attraction mediated by electrons.
\end{abstract}

\maketitle

%\preprint{APS/123-QED}



In this Letter we consider a hybrid Bose-Fermi system consisting of a
spatially separated but interacting Bose-Einstein condensate (BEC) and Fermi
sea. Such systems have been discussed in relation to cold atom gases and
shown to be extremely rich in fundamental effects including the BCS-BEC\
crossover\cite{Girardeau}, Tonks- Girardeau gases\cite{Tonks}, Cooper
pairing and superconductivity\cite{Stoof}. Hybrid systems remain poorely
studied experimentally as realisation of spatially separated but interacting
Fermi sea and Bose-Einstein condensate of cold atoms is a non-trivial task.
Recently, the potential of semiconductor coupled quantum wells for
realization of BEC\ of cold excitons has been revealed \cite{Butov1}.
Spatially indirect excitons were widely studied both experimentally and
theoretically in recent years (see Ref.\onlinecite{Butov} for a review).
They may strongly interact with remote free electrons due to their dipole
moments. This is why realization of a hybrid Bose-Fermi system in a
semiconductor structure containing $n$-doped quantum wells (QWs), where the
two-dimensional electron gas (2DEG)\ is created, and undoped coupled QWs,
where the exciton BEC\ is induced, seems realistic. Recently, a proposal has
been made for realization of exciton-mediated superconductivity in
microcavity structures \cite{PolaritonSupercond}.

Here we study the effect of interaction of a BEC\ of spatially indirect
excitons with a 2DEG on the energy spectrum of excitations of the exciton
BEC. It is well established that in the absence of free electrons, the
Bogoliubov-like spectrum showing linear dispersion near the ground state
and parabolic dispersion at larger wave-vectors is formed due to
exciton-exciton repulsion. We show that due to their scattering with free
electrons, excitons with non-zero momenta attract each other, which
leads to formation of a roton-like minimum\cite{Landau} in the exciton
dispersion. The depth of the minimum increases with increase of the exciton
concentration, and consequently the roton gap decreases. In this regime, the
critical temperature of the Berezinsky-Kosterlitz-Thouless (BKT) transition for
excitons\cite{Posazhennikova} decreases. Eventually, the roton energy goes
below the ground state energy, and the roton gap collapses, leading to the
collapse of the exciton BEC.

\begin{figure}[tbp]
\includegraphics[width=1.0\linewidth]{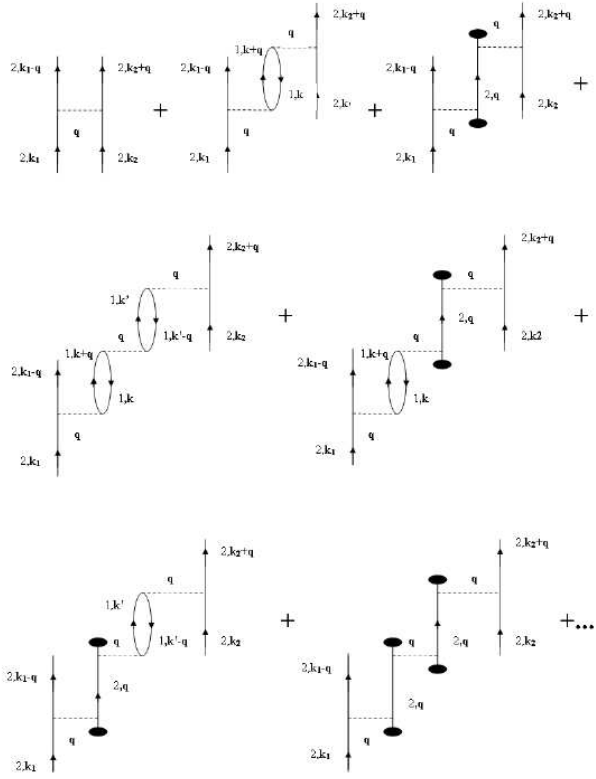}
\caption{RPA diagrammatic representation of the screened interaction between
excitons and electrons. Index 1 corresponds to electrons, index 2 to excitons.
Black dots correspond to the exciton BEC}
\label{FeynmannExcitons}
\end{figure}

Consider a system of three parallel semiconductor quantum wells (QWs), one
of which contains a free electron (or hole) gas, and two others containing
a BEC of spatially indirect excitons. The effective interaction between
electrons and excitons can be represented in diagrammatic form as is
shown in Fig.\ref{FeynmannExcitons}. This figure shows the random phase
approximation (RPA) diagrams for the electronic system and for the virtual
excitations of the exciton BEC. The first diagram represents the direct
interaction of two excitons and all subsequent ones correspond to the
processes involving the virtual excitation of the Fermi sea or/and excitonic
BEC at intermediate stages. For example, the second diagram corresponds to the
process of exciton interaction with the electron Fermi-sea creating a
virtual electron- hole pair in it, which afterwards disappears due to its
interaction with another exciton. Diagram 3 corresponds to the proccess
of exciton interaction with the condensate creating its virtual excitation,
which then disappears due to the interaction with another exciton. Other
diagrams correspond to higher- order processes where several virtual
excitations are created at intermediate stages.

Using the standard rules for the evaluation of Feynman diagrams \cite%
{Zagoskin}, one can obtain the matrix elements of the effective screened
interaction:
\begin{eqnarray}
V_{12}^{eff} &=&V_{12}[1+\left( V_{11}\Pi _{1}+V_{22}\Pi _{2}\right) + \\
&&+\left( V_{11}^{2}\Pi _{1}^{2}+V_{22}^{2}\Pi _{2}^{2}+V_{11}V_{22}\Pi
_{1}\Pi _{2}+V_{12}^{2}\Pi _{1}\Pi _{2}\right) +...],  \nonumber  \label{V12}
\\
V_{11}^{eff} &=&V_{11}+\left( V_{11}^{2}\Pi _{1}+V_{12}^{2}\Pi _{2}\right) +
\\
&&+\left( V_{11}^{3}\Pi _{1}^{2}+2V_{11}V_{12}^{2}\Pi _{1}\Pi
_{2}+V_{12}^{2}V_{22}\Pi _{2}^{2}\right) +...,  \nonumber  \label{V11} \\
V_{22}^{eff} &=&V_{22}+\left( V_{12}^{2}\Pi _{1}+V_{22}^{2}\Pi _{2}\right) +
\\
&&+\left( V_{22}^{3}\Pi _{2}^{2}+2V_{22}V_{12}^{2}\Pi _{1}\Pi
_{2}+V_{12}^{2}V_{11}\Pi _{1}^{2}\right) +...  \nonumber  \label{V22}
\end{eqnarray}%
where $V_{11}(\mathbf{q})$ is the matrix element of the unscreened
interaction between electrons, $V_{22}(\mathbf{q})$ is the matrix element of
the unscreened interaction between excitons, $V_{12}(\mathbf{q})$ is the
matrix element of the unscreened interaction between electrons and excitons.
$\Pi _{j}=\Pi _{j}(\mathbf{q},\omega )$ are polarization operators. For the
electron system one can obtain \cite{Koch}
\begin{equation}
\Pi _{1}(\mathbf{q},\omega )=\sum_{\mathbf{k}}\frac{f_{\mathbf{k-q}}-f_{%
\mathbf{k}}}{\hbar (\omega +i\delta +E_{\mathbf{k-q}}^{el}-E_{\mathbf{k}%
}^{el})},
\label{Pi1}
\end{equation}
and for the condensate \cite{Zagoskin}
\begin{equation}
\Pi _{2}(\mathbf{q},\omega )=N_{0}G_{0}^{ex}(\mathbf{q},\omega )=\frac{%
2N_{0}E_{\mathbf{q}}^{ex}}{(\hbar \omega )^{2}-(E_{\mathbf{q}}^{ex})^{2}}
\label{Pi2}
\end{equation}%
where $N_{0}$ is the occupation number of the condensate, $E_{\mathbf{q}%
}^{ex}$ and $E_{\mathbf{q}}^{el}$ are dispersions of the bare excitons and
electrons, respectively (parabolic, in the effective mass approximation), $%
f_{\mathbf{q}}$ is the Fermi distribution. The electron-electron interaction
is described by a standard 2D Coulomb potential,
\begin{equation}
V_{11}=\frac{e^{2}}{2\epsilon _{0}\epsilon A}\cdot \frac{1}{q}
\end{equation}%
with $\epsilon $ being a dielectric constant of the media and $A$ being the
sample area. The matrix element of exciton-exciton interaction $V_{22}$ can
be and estimated as \cite{Tassone,Ciuti}
\begin{equation}
V_{22}(q)\approx \frac{6E_{B}a_{B}^{2}}{A}
\end{equation}%
with $E_{B}$ and $a_{B}$ being the exciton binding energy and Bohr radius,
respectively. We neglect the $q$-dependent contribution due to dipole-dipole
interaction of excitons, which is essential for the problem of quantum
diffusion of indirect excitons  \cite{Ivanov} but is exactly 0 for  $q=0$
and small compared to exciton-electron interaction for small $q$ in the case
of a uniform exciton BEC, which we consider here.\ The expression for the
matrix element of electron-exciton interaction reads (the details of
calculation can be found in e.g. Ref.\onlinecite{Ramon}):
\begin{eqnarray}
V_{12}(q) &=& \\
&=&\frac{ede^{-qL}}{2\epsilon _{0}\epsilon A}\left\{ \frac{\beta _{e}}{\left[
1+(\frac{\beta _{e}qa_{B}}{2})^{2}\right] ^{3/2}}+\frac{\beta _{h}}{\left[
1+(\frac{\beta _{h}qa_{B}}{2})^{2}\right] ^{3/2}}\right\} +  \nonumber \\
&&+\frac{e^{2}e^{-qL}}{2\epsilon _{0}\epsilon qA}\left\{ \frac{1}{\left[ 1+(%
\frac{\beta _{e}qa_{B}}{2})^{2}\right] ^{3/2}}-\frac{1}{\left[ 1+(\frac{%
\beta _{h}qa_{B}}{2})^{2}\right] ^{3/2}}\right\}  \nonumber
\end{eqnarray}%
where $\beta _{e,h}=m_{e,h}/(m_{e}+m_{h}),$ with $m_{e,h}$ being the
effective masses of electron and hole, $d$ being the dipole moment of the
exciton in normal to the QW plane direction. In order to interact
efficiently with electrons, the excitons must have a significant dipole
moment ($d\sim a_{B})$ which is readily achieved in biased coupled QW
structures where one of the wells confines electrons and another one
confines holes \cite{Butov}.

The formula for the effective interaction can be rewritten in a compact
matrix form
\begin{equation}
\mathbf{V}^{eff}=\mathbf{V}\cdot \left( 1-\Pi \mathbf{V}\right) ^{-1}
\label{Veff}
\end{equation}%
where $\mathbf{V}^{eff},\mathbf{V}$ and $\Pi $ are $2\times 2$ matrices
given above. The polarization matrix $\Pi $ is diagonal, with the diagonal
elements given by expressions \ref{Pi1},\ref{Pi2}.

Expressions \ref{Veff} fully determine all renormalized interactions in the
electron-exciton system and describe simultaneously the screening effects,
the Bogoliubov renormalization of the dispersion of the excitations of the
condensate, the bogolon-mediated pairing of electrons responsible for
exciton mediated superconductivity \cite{PolaritonSupercond}, and effective
attraction between excitons due to their scattering with electrons. The
matrix element of effective exciton-exciton interaction taking into account
all these effects reads
\begin{eqnarray}
V^{eff}_{ex-ex}(q,\omega)= \\
=\frac{V_{22}+\frac{V_{12}^2(q)\Pi_1(q,\omega)}{1-V_{11}(q)\Pi_1(q,\omega)}%
} {(\hbar\omega)^2-(E^{ex}(q))^2-2N_0\left[V_{22}+\frac{V_{12}^2(q)\Pi_1(q,%
\omega)}{1-V_{11}(q)\Pi_1(q,\omega)}\right]E^{ex}(q)}  \nonumber
\end{eqnarray}

The first term in the numerator corresponds to the direct repulsive
exciton-exciton interaction, while the second term describes the effective
interaction between excitons due to the virtual excitations in the
electronic system. If one neglects electron-electron interactions and
polarizability assuming $V_{11}(q)=0,\Pi (q,\omega )=0,$ for the effective
interactions one recovers the result of Refs.\cite{Heiselberg,Stoof}.

The poles of the effective potential determine the dispersions of the collective
modes of the system, given by the equation:
\begin{equation}
(\hbar \omega )^{2}=(E^{ex}(q))^{2}+2N_{0}\tilde{V}_{22}\left( q\right)
E^{ex}(q),  \label{disp}
\end{equation}
where
\begin{equation}
\tilde{V}_{22}\left( q\right) =V_{22}+V_{12}^{2}(q)\Pi _{1}(q,\omega
)/[1-V_{11}(q)\Pi _{1}(q,\omega )].  \label{effpot}
\end{equation}

If the excitonic and electronic systems are uncoupled, $V_{12}=0$, two
independent collective modes coexist: the Bogoliubov excitations of the
condensate (bogolons) whose dispersion is given by $\hbar \omega _{B}(q)=%
\sqrt{E^{ex}(q)[E^{ex}(q)+2V_{22}N_{0}]},$ and the plasmon mode whose
dispersion can be obtained from the transcendental equation $1-V_{11}(q)\Pi
_{1}(q,\omega )=0$. On the other hand, if $V_{12}\neq 0$ and $\Pi _{1}\neq 0$
dispersions of bogolons and plasmonic excitations are coupled.

In this regime, the roton minimum may appear in the energy spectrum of the
exciton condensate. One can see that $\tilde{V}_{22}\left( 0\right) =V_{22}$
as $V_{11}(0)=\infty .$ Assuming $\omega \approx 0$ and using for $\Pi _{1}$
the static approximation, $\Pi _{1}\approx A\frac{m_{el}}{\pi \hbar ^{2}}%
\left( e^{-\pi \hbar ^{2}n_{el}/k_{B}Tm_{el}}-1\right) <0$ where $m_{el}$
and $n_{el}$ are the effective masses and 2D concentration of the electrons,
respectively \cite{Koch}, one can easily see that the second term in the
right part of Eq. (\ref{effpot}) is negative, and $\tilde{V}_{22}$ is a
decreasing function of $q$, which is responsible for the appearance of the
roton minimum.

\begin{figure}[tbp]
\includegraphics[width=1.0\linewidth]{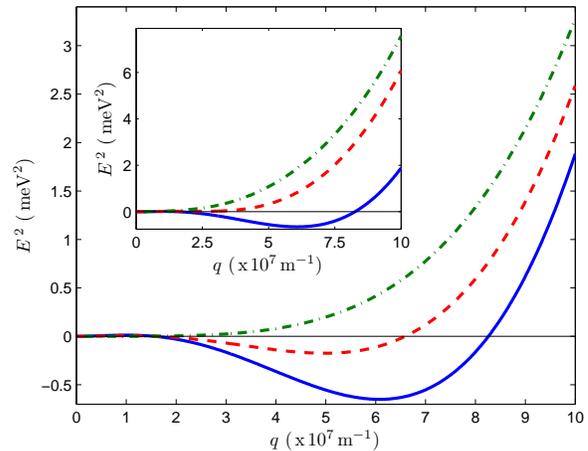}
\caption{Dispersion of the elementary excitations of the condensate showing
roton minimum. Plotted for parameters of a coupled QW structure studied in
Ref. (\protect\cite{Butov1}) assuming that a third \emph{n}-doped QW is
grown at a distance, $L$ from the coupled QW structure, the dipole
separation is $l=12\mathrm{nm}$ and $n_{el}=4\mathrm{x}10^{12}\mathrm{cm}%
^{-2}$. In the main plot $L=12 \mathrm{nm}$ and $n_{s}$ varies as $%
1,0.5,0.01 \mathrm{x}10^{11}\mathrm{cm}^{-2}$(solid blue, dashed red and
dot-dashed green respectively). In the inset $n_s = 1 \mathrm{x} 10^{11}
\mathrm{cm}^{-2}$ and $L$ varies as $12, 25, 55 \mathrm{nm} $ (solid blue,
dashed red and dot-dashed green respectively).}
\label{Dispersion}
\end{figure}

The position and depth of this minimum are dependent on the strength of
the exciton-electron interaction as Figure \ref{Dispersion} shows.

The mimimum becomes deeper if the distance between the QW containing the
electron gas and the exciton BEC\ decreases. It also deepens with
increase of the exciton concentration in the condensate. At some critical
concentration the energy of the roton mimimum equals the energy of the
condensate at $q=0$, so that the roton gap collapses. This manifests itself as a
phase transition in the system: beyond this point the exciton BEC becomes unstable
and eventually collapses due to uncontrollable escape of excitons towards
the roton minimum. The physical reason for suppression of the exciton BEC is
attraction between excitons induced by their interaction with free electrons.

\begin{figure}[tbp]
\includegraphics[width=1.0\linewidth]{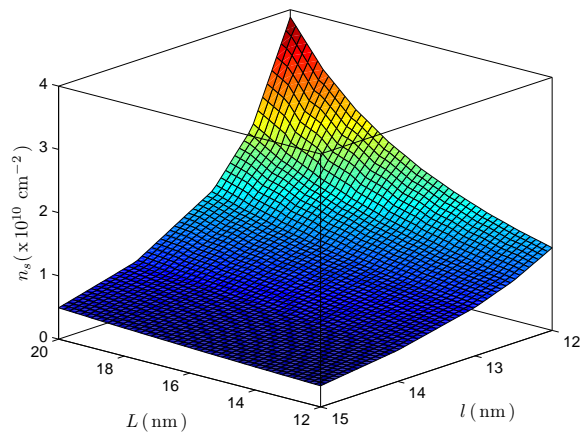}
\caption{Phase diagram of the system - the exciton BEC is unstable above the
phase boundary surface, and stable below. Plotted for the same structure as
in the previous figure.}
\label{PhaseDiagram}
\end{figure}

Figure \ref{PhaseDiagram} shows the phase diagram of the transition between
the exciton BEC\ and classical condensation. One can see that the
collapse of the BEC\ may be achieved at excitons concentrations much below
the Mott density provided that the distance between excitons and electrons
is small enough.

The calculations above have been done assuming zero temperature $T$. If the
superfluid phase exists at $T=0$, it may be also found at higher
temperatures up to the critical temperature which can be found following
Kosterlitz and Nelson \cite{KosterlitzNelson}:
\begin{equation}
T_{BKT}=\frac{\pi \hbar ^{2}n_{s}(T_{BKT})}{2M}  \label{TKT}
\end{equation}%
where $M$ is the exciton mass, $n_{s}$ is a superfluid density. As $n_{s}$
is a function of temperature, the above expression represents a
transcedental equation for $T_{BKT}$.

$n_{s}$ can be found as
\begin{equation}
n_{s}=n-n_{n}(T_{BKT})  \label{density}
\end{equation}%
where $n=N_{0}/A$ is a total 2D concentration of excitons (it equals $%
n_{s}$ at zero temperature), $n_{n}$ is a normal fraction concentration
which can be estimated as in \cite{Berman,Posazhennikova,Malpuech},
\begin{equation}
n_{n}(T,n)=\frac{\hbar ^{2}}{2\pi Mk_{B}T}\int_{0}^{\infty }\frac{%
q^{3}e^{\hbar \omega (q)/k_{B}T}}{(e^{\hbar \omega (q)/k_{B}T}-1)^{2}}dq.
\label{nn}
\end{equation}

The critical temperature $T_{BKT}$ is influenced by the exciton-electron
interaction as it depends on the modified dispersion of the collective modes %
\ref{disp}. Figure 4 shows $T_{BKT}$ for our
system as a function of $L$ and $n$. One can see that at the large $L$
corresponding to a weak exciton-electron coupling, $T_{BKT}$ is a linear
function of the exciton concentration as should be expected for a
conventional gas of interacting bosons. On the other hand, for small $L$
where the exciton-electron interaction is important, the
critical temperature behaves non-monotonously as a function of $n$: it
initially increases, then decreases at larger $n$. This manifests the
gradual changes in the strength and sign of exciton-exciton interactions:
while at low concentrations the excitons repel each other, at higher
concentrations they start attracting each other at large $q$.

\begin{figure}[tbp]
\includegraphics[width=1.0\linewidth]{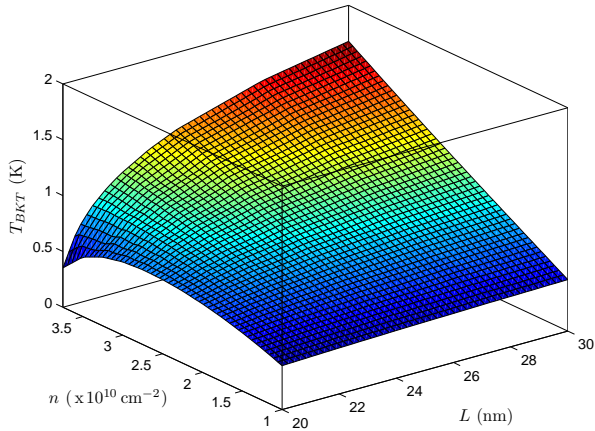}
\caption{Dependence of the Berezinsky-Kosterlitz-Thouless critical
temperature on concentration and QW separation. Plotted for parameters as in
previous figures.}
\label{TKT}
\end{figure}

In conclusion, we have analysed the spectrum of superfluid excitations in a
hybrid exciton-electron system and found the roton minimum whose shape and
depth depend on the exciton concentration and the spatial separation between
excitons and electrons. The electron induced exciton-exciton attraction
leads to a decrease of the BKT\ tranistion temperature and can eventually
destroy the exciton BEC.

The authors thank H. Ouerdane and M. Portnoi for useful discussions. I.A.S.
acknowledges support from RANNIS "Center of excellence in polaritonics".
A.K. acknowledges the E.U. IRSES projects "Robocon" and "Polalas".

\end{document}